\documentclass[10pt,twocolumn,letterpaper]{article}

\usepackage{iccv}
\usepackage{times}
\usepackage{epsfig}
\usepackage{graphicx}
\usepackage{amsmath}
\usepackage{amssymb}
\usepackage{multirow}

\usepackage{tabularx}

\usepackage[labelformat=simple]{subcaption}
\usepackage{capt-of}
\usepackage{booktabs}

\usepackage[pagebackref=true,breaklinks=true,letterpaper=true,colorlinks,bookmarks=false]{hyperref}

\iccvfinalcopy 

\def\iccvPaperID{2634} 

\ificcvfinal\fi

\usepackage{authblk}

\makeatletter
\def\ps@myheadings{%
    \let\@oddfoot\@empty\let\@evenfoot\@empty
    \def\@evenhead{\thepage\hfil\slshape\leftmark}%
    \def\@oddhead{{\slshape\rightmark}\hfil\thepage}%
    \let\@mkboth\@gobbletwo
    \let\sectionmark\@gobble
    \let\subsectionmark\@gobble
    }
\renewcommand\maketitle{\par
  \begingroup
    \renewcommand\thefootnote{\@fnsymbol\c@footnote}%
    \def\@makefnmark{\rlap{\@textsuperscript{\normalfont\@thefnmark}}}%
    \long\def\@makefntext##1{\parindent 1em\noindent
            \hb@xt@1.8em{%
                \hss\@textsuperscript{\normalfont\@thefnmark}}##1}%
    \if@twocolumn
      \ifnum \col@number=\@ne
        \@maketitle
      \else
        \twocolumn[\@maketitle]%
      \fi
    \else
      \newpage
      \global\@topnum\z@   
      \@maketitle
    \fi
    \thispagestyle{plain}\@thanks
  \endgroup
  \setcounter{footnote}{0}%
}
\makeatother
\begin{document}

\title{Unsupervised Microvascular Image Segmentation\\ Using an Active Contours Mimicking Neural Network\vspace{-.8cm}}

\author[1]{Shir Gur}
\author[1,2]{Lior Wolf}
\author[3,4]{Lior Golgher}
\author[3,4]{Pablo Blinder}
\affil[1]{The School of Computer Sceince, Tel Aviv University} 
\affil[2]{Facebook AI Research}
\affil[3]{School of Neurobiology, Biochemistry \& Biophysics, George S. Wise Faculty of Life Sciences, Tel Aviv Uni.}
\affil[4]{Sagol School of Neuroscience, Tel-Aviv University}

\maketitle
\thispagestyle{empty}

\begin{abstract}
   The task of blood vessel segmentation in microscopy images is crucial for many diagnostic and research applications. However, vessels can look vastly different, depending on the transient imaging conditions, and collecting data for supervised training is laborious.
  
   We present a novel deep learning method for unsupervised segmentation of blood vessels. The method is inspired by the field of active contours and we introduce a new loss term, which is based on the morphological Active Contours Without Edges (ACWE) optimization method. The role of the morphological operators is played by novel pooling layers that are incorporated to the network's architecture.
   
   We demonstrate the challenges that are faced by previous supervised learning solutions, when the imaging conditions shift. Our unsupervised method is able to outperform such previous methods in both the labeled dataset, and when applied to similar but different datasets. Our code, as well as efficient pytorch reimplementations of the baseline methods VesselNN and DeepVess is available on GitHub \url{https://github.com/shirgur/UMIS}
\end{abstract}

\section{Introduction}

The field of microscopic imaging of blood vessels is evolving rapidly.
For example, several emerging techniques allow rapid volumetric imaging of vascular dynamics in optically clear~\cite{chhabria2018effect} and turbid~\cite{kong2015continuous,har2018pysight} living brains. Such technologies can lead to breakthroughs, both in understanding neuro-vascular interactions in the neocortex and in fine-grained medical diagnosis. However, this potential can only materialize when automatic segmentation algorithms capable of fast and accurate estimation of vascular dynamics from large volumetric movies become available.

In this work, we propose a novel unsupervised algorithm for blood vessel segmentation. The need for an unsupervised technique arises from the limitations of the currently available training data. These limitations include: (i) the datasets are limited in size, due to the amount of expert labor required, (ii) the datasets do not represent the very high variability that exists in imaging conditions between microscopes, between imaged samples, and along the same experiment. (iii) High-throughput imaging modalities\cite{chhabria2018effect, har2018pysight, kong2015continuous} necessitate faster segmentation algorithms that do not rely on human curation.

Since blood vessels and microvessels have a well-defined tree shape, and are typically brighter than their surroundings (due to the contrast agent), classical (non-learning) algorithms can be used to tackle the problem. One relatively successful method is the Active Contours Without Edges (ACWE) method. However, the method is not accurate enough to outperform learning-based techniques.

We build an unsupervised deep learning technique that is inspired by the morphological ACWE method. The main loss that we employ is motivated by the energy function of ACWE. In addition, we introduce layers that implement the morphological operators that ACWE employs. The result is an unsupervised method that outperforms the supervised methods that were applied in this field.

As an unsupervised method, our models are less likely than the supervised methods to overfit the specific training dataset. We are able to present evidence that applied across datasets (trained on one dataset, applied to another), the new method has an even larger performance gap from the supervised deep learning methods in the literature.

\section{Related Work}

\noindent\textbf{Microvasscular segmentation\quad} 
Multi-photon laser scanning microscopy (MPLSM) allows minimally-invasive investigation of turbid samples, such as living brains, with sub-cellular resolutions~\cite{Denk1990TwophotonLS,Zipfel2003NonlinearMM,Hoover2013AdvancesIM,urban2017understanding}. 
Rapid volumetric movies can be reconstructed by steering the laser beam across the volume and attributing the collected photons to the illuminated voxels~\cite{har2018pysight,kong2015continuous}, thereby utilizing photons that underwent multiple scatterings through the brain, from the focal volume to the detector~\cite{Hoover2013AdvancesIM}.

Blood vessel segmentation has been researched in several domains, such as 3D CT and MRI~\cite{Lesage2009ARO, Fischl2002WholeBS,Bouillot20183DPC}, or 2D retinal blood vessels, see~\cite{Ricci2007RetinalBV,Liskowski2016SegmentingRB, Fu2016RetinalVS} to name but a few. 
More relevant to this paper, are the works of Tsai \etal~\cite{Tsai2009CorrelationsON} for neurons and microvascular segmentation and Mille \etal~\cite{Mille2009DeformableTM} for 2D and 3D branching structures extraction.

Neural network models for vessel segmentation are faster and more accurate than classical methods.
Techniques for retinal blood vessel segmentation, such as~\cite{Wu2016DeepVT,Fu2016RetinalVS,Maninis2016DeepRI} use CNNs and RNNs to perform supervised patch based segmentation of 2D images. In the images domain of two-photons microscopy, Cicek \etal~\cite{cciccek20163d} proposed a 3D-Unet for vascular segmentation, Teikari \etal~\cite{teikari2016deep} introduced VesselNN, which is a 2D-3D network architecture for 3D segmentation. Haft-Javaherian \etal~\cite{haft2018deep} recently proposed DeepVess, which stands as the current state-of-the-art.  Additional  algorithms for automated volumetric segmentation of microvasculature have been recently described in \cite{di2018whole, damseh2018automatic, todorov2019automated}.

\medskip
\noindent\textbf{Active contours\quad} 
First introduced by Kass \etal~\cite{kass1988snakes}, active contours or snakes, are energy-minimizing methods guided by external constraint forces that pull the snake or contour towards features, such as lines and edges. These methods typically require a user-specified initial contour to start with. Variants of this method have increased the robustness and extended the applicability to new domains. A geometric model for active contours was introduced by Caselles \etal~\cite{caselles1993geometric} and Yezzi \etal~\cite{Yezzi1997AGS} with application to CT and MRI images. Kichenassamy \etal~\cite{kichenassamy1995gradient} incorporated gradients flows into the method. Geodesic Active Contour (GAC) by Caselles \etal~\cite{caselles1997geodesic} deform according to an intrinsic geometric measures of the image, induced by image features such as border. 

In the field of optimization based active contours, the ACWE work of Chan and Vese~\cite{chan2001active} and the work of Marquez-Neila \etal on morphological active contours~\cite{marquez2014morphological} are the most relevant to our work. In ACWE, the borders do not have to be well-defined by gradients, and the minimization of the energy functional can be seen as a minimal partition problem. In morphological active contours, the partial differential equations (PDE) used for calculating the curve evolution over time, are replaced by more computationally efficient morphological operators. We give a brief review of morphological ACWE in Sec.~\ref{sec:acwe}.

In the learning-based field, active contours did not fully integrate as a learning concept, but as a method that can be better optimized by a neural network. Rupprecht~\etal~\cite{rupprecht2016deep} for example, trained a class-specific CNN, which predicts a vector pointing from the evolving contour towards the closest point on the boundary of the object of interest. Similarly, Marcos \etal~\cite{marcos2018learning} learn an active contour model parameterizations per instance, using a CNN. 
Other level-sets methods, such as~\cite{hu2017deep,kim2019cnn}, use an additional loss that minimizes the level-set functional, while we use the Euler-Lagrange in Eq.~\ref{eq:EL} and a loss inspired by the ACWE algorithm.
In all methods, supervised learning is used.

We introduce the first full integration between an active contour model and a learning-based model, where we use the image attachment term of the ACWE as a loss, and the morphological curvature operators as a new type of network layer. 
Together, we obtain an unsupervised learning algorithm for performing vascular segmentation.

\section{Active Contours Without Edges}
\label{sec:acwe}
In the following section, we present a brief review of the morphological ACWE method, since it motivates our learning based method. For a detailed explanation, we kindly refer the reader to~\cite{marquez2014morphological}.

Let $\mathcal{C} : \mathbb{R}^+ \times [0,1] \rightarrow \mathbb{R}^3$ be the parametrized 3D curve over time. We start by describing the contour evolution by a partial differential equation (PDE). A differential operator $L$ defines the curve evolution over time with the PDE $\mathcal{C}_t=L(\mathcal{C})$, and $L$ can be rewritten as $L(\mathcal{C}) = \mathcal{F} \cdot \mathcal{N}$, which is the product between the normal $\mathcal{N}$ to the curve, and the scalar field $\mathcal{F}$, which determines the velocity of evolution for each point in the curve.

We consider a level-set representation of $\mathcal{C}$, defined as $u : \mathbb{R}^+ \times \mathbb{R}^3 \rightarrow \mathbb{R}$ such that $\mathcal{C}(t) = \{(x,y,z):u(t,(x,y,z))=0\}$, to express the evolution of the curve as~\cite{osher1988fronts,kimmel2003numerical}:
\begin{equation}
    \frac{\partial u}{\partial t} = \mathcal{F} \cdot |\nabla u|
\end{equation}
where $\mathcal{F}$ can be defined as the normal to the curve, \ie $\mathcal{F} \in \{1, -1\}$, or as the intrinsic heat equation~\cite{guichard2004contrast}, \ie $\mathcal{F} = \mathcal{K}$, where $\mathcal{K}$ is the Euclidean curvature of $\mathcal{C}$.

\begin{figure}[t]
    \centering
    \includegraphics[width=.9\linewidth]{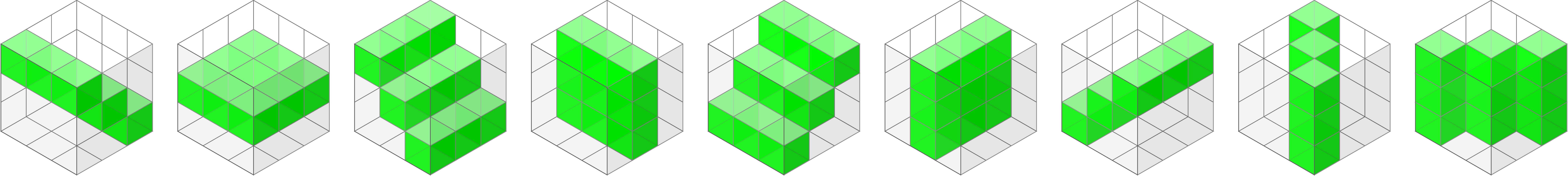}
    \caption{Illustration of the nine 3D structuring elements of $\mathcal{B}$. Also used as masks in the morphological pooling layer.}
    \label{fig:B}
\end{figure}

\smallskip
\noindent\textbf{3D Curvature morphological operator\quad} 
Monotone contrast-invariant and translation-invariant operators are called morphological operators, such as dilation and erosion.
Let $\mathcal{B}$ be a set of structuring elements that uniquely defines a morphological operator, and $h$ is the size of that operator, we now define two morphological operators:
\begin{align}
    SI u(x) = \sup_{B \in \mathcal{B}}\inf_{y \in x+hB} u(y)\\
    IS u(x) = \inf_{B \in \mathcal{B}}\sup_{y \in x+hB} u(y)
\end{align}
where $\mathcal{B}$ contains all hyper-disks of radius 1 centered at the origin, and in our three-dimensional discrete case, the nine rectangular surfaces centered at the origin of a $3\times3\times3$ cube (see Fig.~\ref{fig:B}).

In their study,~\cite{marquez2014morphological} showed that $\mathcal{K}$ can be expressed by a curvature morphological operator defined as $SI \circ IS$.

\smallskip
\noindent\textbf{Morphological ACWE\quad} 
Let $\mathcal{C}$ be the parametrized curve evolution and $I$ be an image, the Morphological ACWE functional to be minimized as defined in the work of~\cite{marquez2014morphological, chan2001active} is as follows:
\begin{equation}
    \begin{split}
        F(c_1, c_2, \mathcal{C}) = &\mu \cdot \text{length}(\mathcal{C}) + v \cdot \text{area}(\text{inside}(\mathcal{C}))\\
        & + \alpha \int_{\text{inside}(\mathcal{C})} ||I(x) - c_1 || dx\\
        & + \beta \int_{\text{outside}(\mathcal{C})} ||I(x) - c_2 || dx
    \end{split}
\end{equation}
where $c_1$ and $c_2$ are the mean of the values inside and outside $\mathcal{C}$, and defined as follows:
\begin{align}
    c_1(\mathcal{C}) = \frac{\int_{\text{inside}(\mathcal{C})} I(x) dx}{\int_{\text{inside}(\mathcal{C})} dx}\label{eq:c1}\\
    c_2(\mathcal{C}) = \frac{\int_{\text{outside}(\mathcal{C})} I(x) dx}{\int_{\text{outside}(\mathcal{C})} dx}\label{eq:c2}
\end{align}
The parameters $\mu$ and $v$ control the two terms of length (or curvature) and area of $\mathcal{C}$, as defined in~\cite{chan2001active}.
The Euler-Lagrange equation for the functional $F$ (see~\cite{marquez2014morphological} for details) is as follows:
\begin{equation}
    \label{eq:EL}
    \begin{split}
    \frac{\partial u}{\partial t} = &|\nabla u| \bigg(\mu \text{div} \bigg( \frac{\nabla u}{|\nabla u|} \bigg) - v\\
    &- \alpha(I-c_1)^2 + \beta(I-c_2)^2 \bigg)
    \end{split}
\end{equation}
where in the experiments of~\cite{marquez2014morphological}, $v$ is set to 0, and the parameters $\alpha$ and $\beta$ are set to 1 and 2, respectively. $\nabla u$ is computed using central differences along the three axis $x$,$y$, and $z$.
The resulting ACWE algorithm is written as follows:
\begin{align}
    \label{eq:AC_energy}
    \Gamma &= |\nabla u|( \alpha (I - c_1)^2 - \beta (I - c_2)^2 )\\
    \label{eq:AC_alg}
    \begin{split}
        u^{n+\frac{1}{2}}(x) &= 
        \begin{cases}
            1 & \text{if }\Gamma < 0\\
            0 & \text{if }\Gamma > 0 \\
            u^{n} & \text{otherwise}
        \end{cases}
    \end{split}\\
    \label{eq:AC_alg2}
    u^{n+1}(x) &= \big((SI \circ IS)^\mu u^{n+\frac{1}{2}}\big) (x)
\end{align}
where $\Gamma$ is the image attachment term, as defined by~\cite{marquez2014morphological}. The superscript is the iteration number.

\section{Method}
\begin{figure*}[t]
    \centering
    \includegraphics[width=\linewidth]{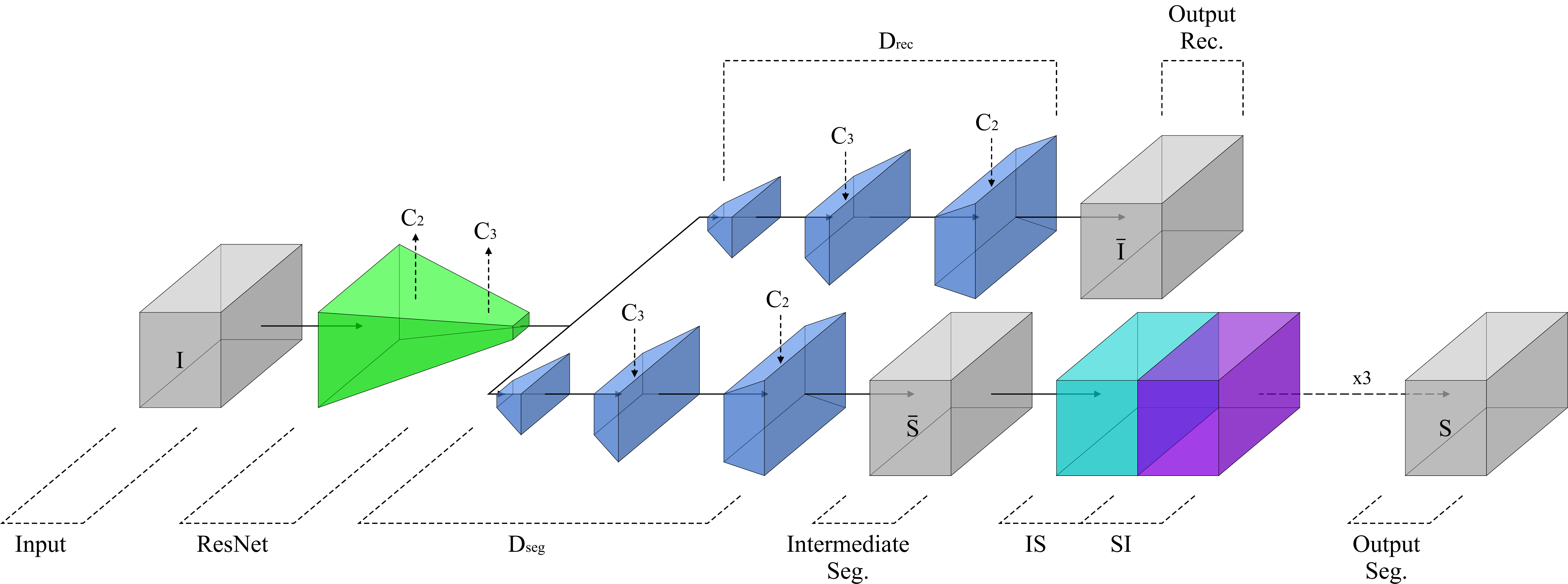}
    \caption{3D segmentation network architecture. $C2$ and $C3$ are intermediate convolution blocks from ResNet, serving as skip-inputs for both $D_{seg}$ and $D_{rec}$.}
    \label{fig:arch}
\end{figure*}
In order to implement the ACWE principle as a network, we turn the iterative energy minimization that occurs in Eq.~\ref{eq:AC_alg} into a loss, and the morphological operations in Eq.~\ref{eq:AC_alg2} into novel morphological layers.

The segmentation network receives an input $I\in [0, 1]^{1 \times k \times m \times n}$, which is a 3D intensity-response input volume, where $k \times m \times n$ are the volume dimensions of a single intensity channel.
The network outputs a segmentation map, $S \in [0, 1]^{1 \times k \times m \times n}$, and thresholding is performed to obtain the final result. To lower the ambiguity of $S$, and the sensitivity to thresholding, we employ a compound loss, which encourages output results to have 0 or 1 values.

\subsection{Network Architecture}
The network's architecture is illustrated in Fig.~\ref{fig:arch} and consists of a main Encoder-Decoder branch with skip connections, denoted as $E$ and $D_{seg}$ (for \textit{segmentation}), respectively, followed by successive operations of Morphological Pooling Layer (Eq.~\ref{eq:morphpool1}-\ref{eq:morphpool2}) for smoothing.
It is trained in an unsupervised manner, using an auxiliary reconstruction loss, provided by an additional decoder $D_{rec}$, which is used only during training, and outputs an estimated input $\bar{I}$. 

The network components are then rewritten as follows:
\begin{align}
    \bar{S}(I) &:= D_{seg}(E(I))\\
    S(I) &:= \underbrace{SI(IS(\dots(SI(IS}_{SI \circ IS\; \mu \textnormal{ times}}(\bar{S}(I))))))\label{eq:M}\\
    \bar{I} &:= D_{rec}(E(I))
\end{align}
where $\bar{S}$ is the segmentation before smoothing, and $S$ is the segmentation mask obtained after applying the morphological pooling layers $SI$ and $IS$ in Eq.~\ref{eq:M} $\mu$ times (the two layers are defined below).

Our Encoder architecture is based on ResNet34~\cite{He2016DeepRL} with 3D convolutions. In addition to the ResNet's output $E(I)$, there are two intermediate layers that are fed directly into the subsequent decoder: $C2$ ($C3$) is the output of the ResNet's 2nd (3rd) block. Each of the two decoders $D_{seg}$ and $D_{rec}$ consists of three Upsampling blocks with skip connections, as detailed in the appendix.

\begin{figure}[t]
    \centering
    \includegraphics[width=\linewidth]{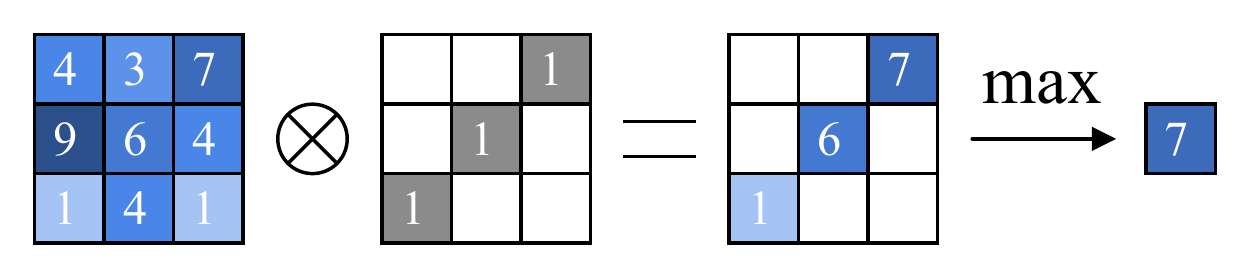}
    \caption{Illustration of 2D Masked Pooling Layer activation on a single window.}
    \label{fig:maskpool}
\end{figure}

\smallskip
\noindent\textbf{Morphological Pooling Layer\quad} 
We have implemented the $IS$ and $SI$ operations as differentiable layers, which are not learned, but play a role in the backpropagation. These layers employ the set of nine structuring elements $\mathcal{B}$, as explained in Sec.~\ref{sec:acwe}, where each element $B \in \mathcal{B}$ is a binary mask of size $3\times 3 \times 3$ as illustrated in Fig.~\ref{fig:B}. The layers first perform masked max pooling $\forall B \in \mathcal{B}$, and then take the {\em maximum} or {\em minimum} across all results, according to the desired operation ($SI$ or $IS$ respectively). Formally:
\begin{align}
    SI(x) &:= \max\limits_{B \in \mathcal{B}} -MaskPool(-x, B)\label{eq:morphpool1}\\
    IS(x) &:= \min\limits_{B \in \mathcal{B}} MaskPool(x, B)\label{eq:morphpool2}\\
    MaskPool(x, B) &:= \max \{x \otimes B\}
\end{align}
The $MaskPool$ function first applies an element-wise multiplication between the mask and the input, denoted by $\otimes$, and then takes the maximum over all locations. Fig.~\ref{fig:maskpool} illustrates the 2D case.

The two types of Morphological Pooling Layers are then used to smooth the intermediate segmentation $\bar{S}$ and obtain the final segmentation mask $S$, according to Eq.~\ref{eq:M}, where the two morphological operations are applied $\mu=3$ times.

\subsection{Training loss}
\label{subsec:loss}
We employ a compound loss:
\begin{equation}
    \begin{split}
        \mathcal{L} = &\lambda_1\mathcal{L}_{AC} + \lambda_2\mathcal{L}_{rank} + \lambda_3\mathcal{L}_{tight} +\\ 
        &\lambda_4\mathcal{L}_{rec} + \lambda_5\mathcal{L}_{MV} + \lambda_6\mathcal{L}_{ME}
    \end{split}
\end{equation}
where $\lambda_1,..,\lambda_6$ are weights. The various terms of the loss function are defined below.

\smallskip
\noindent\textbf{Active contour loss\quad} 
Our main loss term, $\mathcal{L}_{AC}$, is derived from the ACWE algorithm, specifically Eq.~\ref{eq:AC_energy} which determine the segmentation value in Eq.~\ref{eq:AC_alg}.
We first rewrite Eq.~\ref{eq:AC_energy} to fit our network terminology:
\begin{equation}
    \Gamma = \|\nabla \bar{S}\|_1( \alpha (I - c_1)^2 - \beta (I - c_2)^2) \label{eq:gamma}
\end{equation}
where $\nabla \bar{S}$ is the intermediate segmentation derivatives in $x$, $y$, and $z$, computed using the central differences method.
We define $c_1$ and $c_2$ similarly to Eq~\ref{eq:c1}-~\ref{eq:c2} as:
\begin{align}
    c_1 &= \frac{\sum_p I(p)S(p)}{\sum_p S(p)} \label{eq:c1_net}\\
    c_2 &= \frac{\sum_p I(p)(1-S(p))}{\sum_p 1-S(p)}\label{eq:c2_net}
\end{align}
Following~\cite{marquez2014morphological}, we set $\alpha = 1$ and $\beta = 2$ for all experiments.

As can be seen from Eq.~\ref{eq:AC_alg}, when $\Gamma<0$, we need to encourage the segmentation to output 1, and conversely, output 0, if $\Gamma>0$. Therefore, we penalize elements of $S$ for which this is not the case. This is done by the following loss, which is computed for each point $p$ in the volume:
\begin{equation}
    \label{eq:loss_ac}
    \begin{split}
        \mathcal{L}_{AC}(p) = 
        \begin{cases}
            \exp(\Gamma(p) S(p)) & 
            \begin{split}
                \text{if }\Gamma(p) <= 0
            \end{split}\\
             \exp(-\Gamma(p)  (1 - S(p))) & 
            \begin{split}
                \text{if }\Gamma(p) > 0
            \end{split}
        \end{cases}
    \end{split}
\end{equation}

The volume loss is $\mathcal{L}_{AC} = \mathbb{E}_p \mathcal{L}_{AC}(p)$. The loss is high, if the exponent is applied to a value that is close to zero. This happens if the term $\Gamma(p)$ is negative and $S(p)$ is close to zero, or if $\Gamma(p)$ is positive and $S(p)$ is close to one.

\smallskip
\noindent\textbf{Ranking loss\quad}
$\mathcal{L}_{rank}$ plays two roles in the learning process. Firstly, it enforces $c_1$ to represent the higher values in the input image, segmenting the reflective substance. Secondly, it encourages a larger gap between $c_1$ and $c_2$. 
\begin{equation}
    \mathcal{L}_{rank} = exp(c_2 - c_1) 
\end{equation}

\noindent\textbf{Reconstruction loss\quad}
Since learning is performed without supervision, we further constrain the encoder to preserve the information of the input data by adding the $\mathcal{L}_{rec}$ reconstruction loss:
\begin{equation}
    \mathcal{L}_{rec} = \mathbb{E}_p \big[ (\bar{I}(p) - I(p))^2 + \|\nabla \bar{I}(p)\|_1 \big]
\end{equation}
where $\nabla \bar{I}$ is a regularization (smoothing) term.

\smallskip
\noindent\textbf{Minimal segmentation loss\quad}
To avoid the situation in which the segmentation mask contains the entire image, which seems to be a stable solution when the training data is noisy, we encourage the network to output a tight (or minimal) segmentation:
\begin{equation}
    \mathcal{L}_{tight} = A(S) = \sum_p S(p)
\end{equation}
where $A(S)$ is the area of segmentation $S$.

\smallskip
\noindent\textbf{Disjunctive loss\quad}
We add two additional loss terms to push the segmentation output away from the intermediate values and towards 0 or 1. The first is the Maximal Variance loss:
\begin{equation}
    \mathcal{L}_{MV} = \exp(\mathbb{E}_p[S(p)^2] - \mathbb{E}[S(p)]^2)
\end{equation}
and the second is the Maximal Entropy loss:
\begin{equation}
    \mathcal{L}_{ME} = \mathbb{E}_p[-S(p) \cdot \log(S(p))]
\end{equation}

\subsection{Training the network}

The same settings are used in all experiments and the hyperparameters are fixed to the following values: $\lambda_1=1, \lambda_2=10^{-2}, \lambda_3=\lambda_4=10^{-3}, \lambda_5 = 10^{-3}$ and $\lambda_6=10^{-6}$. An Adam optimizer, with a learning rate of $0.0001$, is used. Training is done on a single NVIDIA Titan-X 12G Pascal GPU, and takes about 20 minutes to train.

The 3D input volume of our network has a size of $32 \times 128 \times 128$, where $32$ is for the $z$ dimension. At train time, we crop the training samples to fit the network's input shape. At test time, a sliding window is applied, with a stride of $8 \times 16 \times 16$. The resulting outputs of the network are then averaged at every 3D location. Reflection padding is applied to the test images before slicing, to support a uniform sampling pattern across the entire volume. 

\subsection{Unsupervised fine tuning}
An unsupervised method has the advantage that it can be applied to the test data without seeing the labels, i.e., one can naturally apply transductive learning and update the network based on the test data. This is especially helpful in the cross-domain scenario.

When we follow this protocol (we mark it explicitly as FT), the network, which was previously pre-trained on the training data, continues to train on the test data. In all of our transductive learning experiments, we use the same learning rate for both training and fine-tuning. For a fair comparison with supervised methods, we perform the additional training for the same length of time, as it takes to evaluate the test set with the relatively fast DeepVess method~\cite{haft2018deep}.

\section{Experiments}

We present three types of experiments. First, we compare our method with supervised literature neural network methods such as DeepVess~\cite{haft2018deep} by Haft-Javaherian \etal and VesselNN~\cite{teikari2016deep} by Teikari \etal, and classic optimization methods, such as the morphological ACWE method presented above. This is done in the supervised setting, where one trains on the training split and tests on the test split of the same dataset. Second, we test the generalization capability of each method in a cross-datasets evaluation, where each method is first trained on the training set of dataset $A$ and then tested on the test set of dataset $B$. Finally, we present qualitative results on a new 4D microvascular intravital dataset of neuronal activity and vascular cross-section volumetric movie.
This dataset suffers from a relatively low Signal to Noise Ratio (SNR), and from sparse images at every time frame.

In our experiments we use two re-implementations of state-of-the-art supervised methods for vessels segmentation, verifying that we are able to obtain the same level of results for these baselines as the original, less efficient, implementations. We also report results for morphological ACWE as well as VIDA~\cite{Tsai2009CorrelationsON} for both datasets, in order to add unsupervised baseline methods.

\begin{table*}[t]
    \centering
    \setlength{\tabcolsep}{1pt} 
    \renewcommand{\arraystretch}{1} 
    \begin{tabular*}{\textwidth}{lcccccc|ccccc}
        \raisebox{6mm}{(a)} &
        \includegraphics[width=0.08\linewidth]{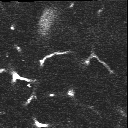} &
        \includegraphics[width=0.08\linewidth]{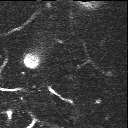} &
        \includegraphics[width=0.08\linewidth]{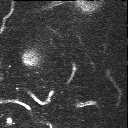} &
        \includegraphics[width=0.08\linewidth]{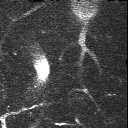} &
        \includegraphics[width=0.08\linewidth]{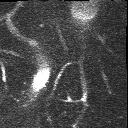} &
        \includegraphics[width=0.08\linewidth]{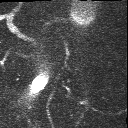} &
        \includegraphics[width=0.08\linewidth]{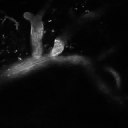} &
        \includegraphics[width=0.08\linewidth]{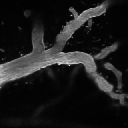} &
        \includegraphics[width=0.08\linewidth]{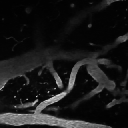} &
        \includegraphics[width=0.08\linewidth]{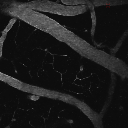} &
        \includegraphics[width=0.08\linewidth]{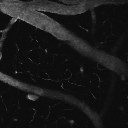}\\
        
        \raisebox{6mm}{(b)} &
        \includegraphics[width=0.08\linewidth]{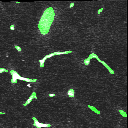} &
        \includegraphics[width=0.08\linewidth]{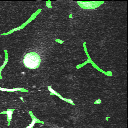} &
        \includegraphics[width=0.08\linewidth]{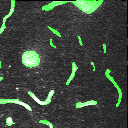} &
        \includegraphics[width=0.08\linewidth]{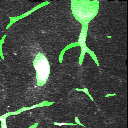} &
        \includegraphics[width=0.08\linewidth]{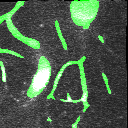} &
        \includegraphics[width=0.08\linewidth]{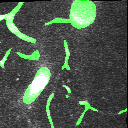} &
        \includegraphics[width=0.08\linewidth]{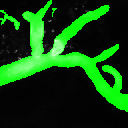} &
        \includegraphics[width=0.08\linewidth]{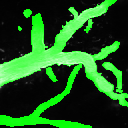} &
        \includegraphics[width=0.08\linewidth]{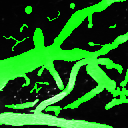} &
        \includegraphics[width=0.08\linewidth]{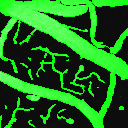} &
        \includegraphics[width=0.08\linewidth]{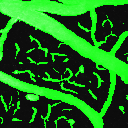} \\
        
        \midrule
        
        \raisebox{6mm}{(c)} &
        \includegraphics[width=0.08\linewidth]{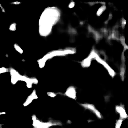} &
        \includegraphics[width=0.08\linewidth]{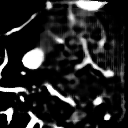} &
        \includegraphics[width=0.08\linewidth]{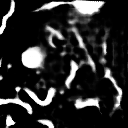} &
        \includegraphics[width=0.08\linewidth]{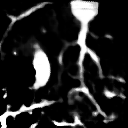} &
        \includegraphics[width=0.08\linewidth]{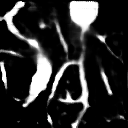} &
        \includegraphics[width=0.08\linewidth]{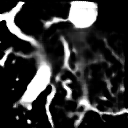} &
        \includegraphics[width=0.08\linewidth]{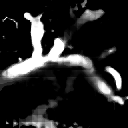} &
        \includegraphics[width=0.08\linewidth]{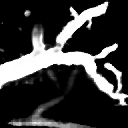} &
        \includegraphics[width=0.08\linewidth]{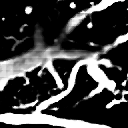} &
        \includegraphics[width=0.08\linewidth]{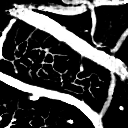} &
        \includegraphics[width=0.08\linewidth]{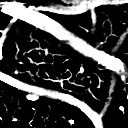} \\
        
        \raisebox{6mm}{(d)} &
        \includegraphics[width=0.08\linewidth]{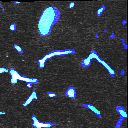} &
        \includegraphics[width=0.08\linewidth]{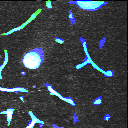} &
        \includegraphics[width=0.08\linewidth]{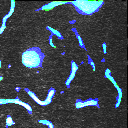} &
        \includegraphics[width=0.08\linewidth]{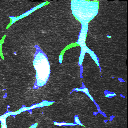} &
        \includegraphics[width=0.08\linewidth]{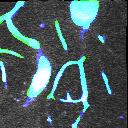} &
        \includegraphics[width=0.08\linewidth]{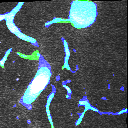} &
        \includegraphics[width=0.08\linewidth]{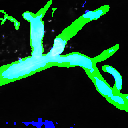} &
        \includegraphics[width=0.08\linewidth]{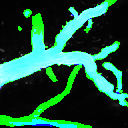} &
        \includegraphics[width=0.08\linewidth]{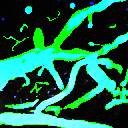} &
        \includegraphics[width=0.08\linewidth]{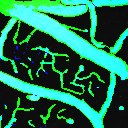} &
        \includegraphics[width=0.08\linewidth]{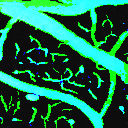} \\
        
        \raisebox{6mm}{(e)} &
        \includegraphics[width=0.08\linewidth]{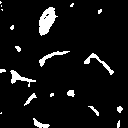} &
        \includegraphics[width=0.08\linewidth]{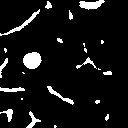} &
        \includegraphics[width=0.08\linewidth]{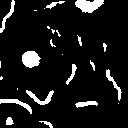} &
        \includegraphics[width=0.08\linewidth]{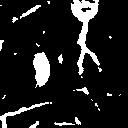} &
        \includegraphics[width=0.08\linewidth]{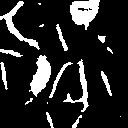} &
        \includegraphics[width=0.08\linewidth]{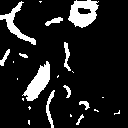} &
        \includegraphics[width=0.08\linewidth]{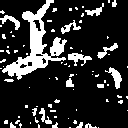} &
        \includegraphics[width=0.08\linewidth]{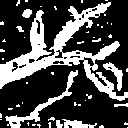} &
        \includegraphics[width=0.08\linewidth]{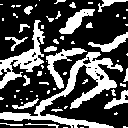} &
        \includegraphics[width=0.08\linewidth]{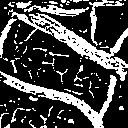} &
        \includegraphics[width=0.08\linewidth]{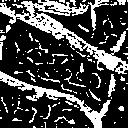} \\
        
        \raisebox{6mm}{(f)} &
        \includegraphics[width=0.08\linewidth]{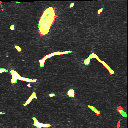} &
        \includegraphics[width=0.08\linewidth]{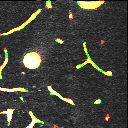} &
        \includegraphics[width=0.08\linewidth]{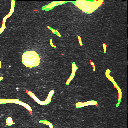} &
        \includegraphics[width=0.08\linewidth]{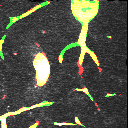} &
        \includegraphics[width=0.08\linewidth]{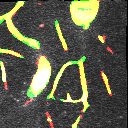} &
        \includegraphics[width=0.08\linewidth]{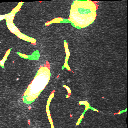} &
        \includegraphics[width=0.08\linewidth]{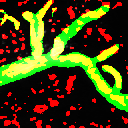} &
        \includegraphics[width=0.08\linewidth]{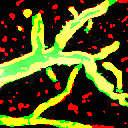} &
        \includegraphics[width=0.08\linewidth]{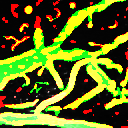} &
        \includegraphics[width=0.08\linewidth]{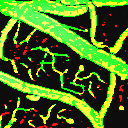} &
        \includegraphics[width=0.08\linewidth]{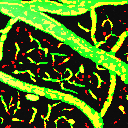} \\
        
        \midrule
        \raisebox{6mm}{(g)} &
        \includegraphics[width=0.08\linewidth]{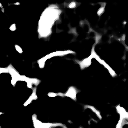} &
        \includegraphics[width=0.08\linewidth]{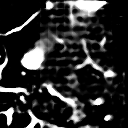} &
        \includegraphics[width=0.08\linewidth]{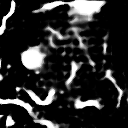} &
        \includegraphics[width=0.08\linewidth]{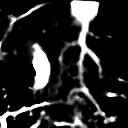} &
        \includegraphics[width=0.08\linewidth]{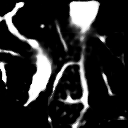} &
        \includegraphics[width=0.08\linewidth]{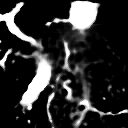} &
        \includegraphics[width=0.08\linewidth]{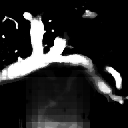} &
        \includegraphics[width=0.08\linewidth]{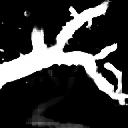} &
        \includegraphics[width=0.08\linewidth]{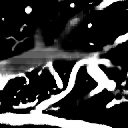} &
        \includegraphics[width=0.08\linewidth]{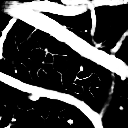} &
        \includegraphics[width=0.08\linewidth]{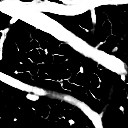} \\
        
        \raisebox{6mm}{(h)} &
        \includegraphics[width=0.08\linewidth]{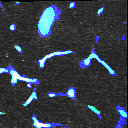} &
        \includegraphics[width=0.08\linewidth]{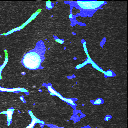} &
        \includegraphics[width=0.08\linewidth]{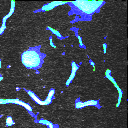} &
        \includegraphics[width=0.08\linewidth]{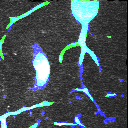} &
        \includegraphics[width=0.08\linewidth]{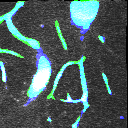} &
        \includegraphics[width=0.08\linewidth]{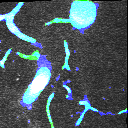} &
        \includegraphics[width=0.08\linewidth]{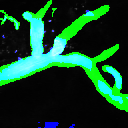} &
        \includegraphics[width=0.08\linewidth]{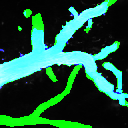} &
        \includegraphics[width=0.08\linewidth]{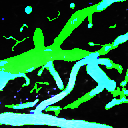} &
        \includegraphics[width=0.08\linewidth]{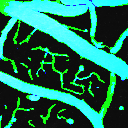} &
        \includegraphics[width=0.08\linewidth]{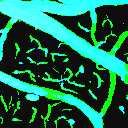} \\
        \raisebox{6mm}{(i)} &
        \includegraphics[width=0.08\linewidth]{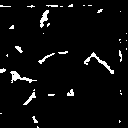} &
        \includegraphics[width=0.08\linewidth]{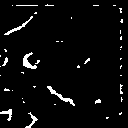} &
        \includegraphics[width=0.08\linewidth]{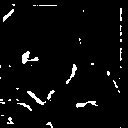} &
        \includegraphics[width=0.08\linewidth]{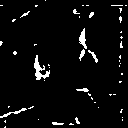} &
        \includegraphics[width=0.08\linewidth]{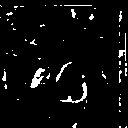} &
        \includegraphics[width=0.08\linewidth]{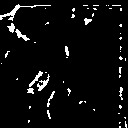} &
        \includegraphics[width=0.08\linewidth]{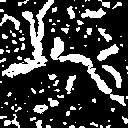} &
        \includegraphics[width=0.08\linewidth]{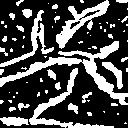} &
        \includegraphics[width=0.08\linewidth]{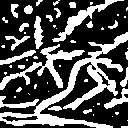} &
        \includegraphics[width=0.08\linewidth]{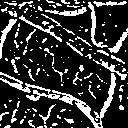} &
        \includegraphics[width=0.08\linewidth]{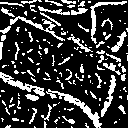} \\
        
        \raisebox{6mm}{(j)} &
        \includegraphics[width=0.08\linewidth]{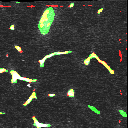} &
        \includegraphics[width=0.08\linewidth]{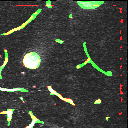} &
        \includegraphics[width=0.08\linewidth]{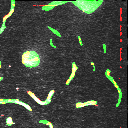} &
        \includegraphics[width=0.08\linewidth]{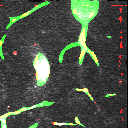} &
        \includegraphics[width=0.08\linewidth]{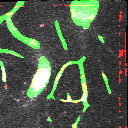} &
        \includegraphics[width=0.08\linewidth]{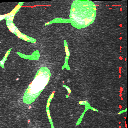} &
        \includegraphics[width=0.08\linewidth]{figures/dv_vnn_dataset/dv_vnn_0_3_ann_dv.png} &
        \includegraphics[width=0.08\linewidth]{figures/dv_vnn_dataset/dv_vnn_0_9_ann_dv.png} &
        \includegraphics[width=0.08\linewidth]{figures/dv_vnn_dataset/dv_vnn_0_17_ann_dv.png} &
        \includegraphics[width=0.08\linewidth]{figures/dv_vnn_dataset/dv_vnn_2_9_ann_dv.png} &
        \includegraphics[width=0.08\linewidth]{figures/dv_vnn_dataset/dv_vnn_2_11_ann_dv.png} \\
        
        \multicolumn{7}{c|}{DeepVess dataset~\cite{haft2018deep}} & \multicolumn{4}{c}{VesselNN dataset~\cite{teikari2016deep}}
        
    \end{tabular*}
    \captionof{figure}{Qualitative segmentation results on the DeepVess dataset (left) and the VesselNN dataset (right). All images are a z slice from either dataset. \textbf{Green} - Expert annotations, \textbf{Blue} - Our segmentation, \textbf{Red} - DeepVess segmentation. \textbf{(a)} Input image \textbf{(b)} Expert annotations. \textbf{(c)-(f)} Results generated by training and testing on the same dataset, \textbf{(c)} Our output \textbf{(d)} Our + annotations, \textbf{(e)} DeepVess output, \textbf{(f)} DeepVess + annotations. \textbf{(g)-(j)} Cross-datasets results where training is preformed on the opposite dataset, \textbf{(g)} Our output \textbf{(h)} Our + annotations, \textbf{(i)} DeepVess output, \textbf{(j)} DeepVess + annotations.}
    \label{fig:qual_results}
\end{table*}

\begin{table*}[t]
    \centering
    \begin{tabular*}{\textwidth}{@{\extracolsep{\fill} } l|cccccccc }
        \toprule
         Algorithm & Supervised & AP & F1 & Sensitivity & Specificity & JI & DICE & mIoU \\
         \midrule
         Cicek \etal~\cite{cciccek20163d} & \checkmark & - & - & 0.700 & 0.982 & 0.594 & 0.726 & -\\
         VesselNN~\cite{teikari2016deep}$^\dagger$ & \checkmark & 0.867 & 0.732 & 0.868 & \textbf{0.989} & 0.777 & 0.732 & 0.765\\
         DeepVess~\cite{haft2018deep}$^\dagger$ & \checkmark & 0.889 & 0.820 & 0.951 & 0.984 & 0.807 & 0.820 & 0.828\\
         \midrule
         VIDA~\cite{Tsai2009CorrelationsON} & & 0.179 & 0.304 & \textbf{0.998} & 0.564 & 0.179 & 0.304 & 0.372 \\
         Morph-ACWE~\cite{marquez2014morphological} & & 0.483 & 0.676 & 0.743 & 0.957 & 0.511 & 0.676 & 0.722 \\
         Ours & & \textbf{0.909} & \textbf{0.829} & 0.992 & 0.986 & \textbf{0.811} & \textbf{0.829} & \textbf{0.838} \\
         \bottomrule
    \end{tabular*}
    \caption{Training and testing on the DeepVess dataset. $^\dagger$represents an improved re-implementation of the method.}
    \label{tab:deepvess}
    \begin{tabular*}{\textwidth}{@{\extracolsep{\fill} } l|cccccccc }
        \toprule
         Algorithm & Supervised & AP & F1 & Sensitivity & Specificity & JI & DICE & mIoU \\
         \midrule
         VesselNN~\cite{teikari2016deep}$^\dagger$ & \checkmark & 0.786 & 0.739 & 0.898 & 0.934 & 0.705 & 0.739 & 0.503\\
         DeepVess~\cite{haft2018deep}$^\dagger$ & \checkmark & 0.804 & 0.757 & 0.901 & 0.986 & 0.713 & 0.757 & 0.629\\
         \midrule
         VIDA~\cite{Tsai2009CorrelationsON} & & 0.591 & 0.589 & 0.452 & 0.984 & 0.418 & 0.589 & 0.624 \\
         Morph-ACWE~\cite{marquez2014morphological} & & 0.505 & 0.528 & 0.367 & \textbf{0.998} & 0.364 & 0.505 & 0.599 \\
         Ours & & \textbf{0.834} & \textbf{0.776} & \textbf{0.923} & 0.929 & \textbf{0.721} & \textbf{0.776} & \textbf{0.760} \\
         \bottomrule
    \end{tabular*}
    \caption{Training and testing on the VesselNN dataset.$^\dagger$represents an improved re-implementation of the method.}
    \label{tab:vesselnn}
    
    \begin{tabular*}{\textwidth}{@{\extracolsep{\fill} } l|cccccccc }
        \toprule
         Algorithm & Transition & AP & F1 & Sensitivity & Specificity & JI & DICE & mIoU \\
         \midrule
         VesselNN~\cite{teikari2016deep}$^\dagger$ & \multirow{4}{*}{\begin{minipage}{1.15in}DeepVess dataset $\rightarrow$ VesselNN dataset\end{minipage}} & 0.528 & 0.188 & 0.110 & \textbf{0.991} & 0.108 & 0.188 & 0.469\\
         DeepVess~\cite{haft2018deep}$^\dagger$ & & 0.664 & 0.562 & 0.469 & 0.960 & 0.395 & 0.562 & 0.604\\
         Ours & & 0.684 & 0.662 & 0.833 & 0.915 & 0.501 & 0.662 & 0.681 \\
         Ours - FT & & \textbf{0.752} & \textbf{0.715} & \textbf{0.901} & 0.940 & \textbf{0.563} & \textbf{0.715} & \textbf{0.727} \\
         \midrule
         VesselNN~\cite{teikari2016deep}$^\dagger$ & \multirow{4}{*}{\begin{minipage}{1.15in}VesselNN dataset $\rightarrow$ DeepVess dataset \end{minipage}} & 0.164 & 0.175 & 0.142 & 0.948 & 0.096 & 0.175 & 0.482 \\
         DeepVess~\cite{haft2018deep}$^\dagger$ & & 0.701 & 0.506 & 0.351 & \textbf{0.996} & 0.505 & 0.506 & 0.637\\
         Ours & & 0.895 & 0.803 & \textbf{0.988} & 0.917 & 0.671 & 0.803 & \textbf{0.821} \\
         Ours - FT & & \textbf{0.902} & \textbf{0.808} & 0.979 & 0.970 & \textbf{0.678} & \textbf{0.808} & 0.815 \\
         \bottomrule
    \end{tabular*}
    \caption{Training on A and testing on B data (A $\rightarrow$ B). FT - Unsupervised fine-tune on A data for $t$ seconds, where $t$ is the time taken for DeepVess to perform evaluation. $^\dagger$represents an improved re-implementation of the method.}
    \label{tab:cross}
\end{table*}

\subsection{Datasets}

The three datasets employed consist of microvascular images produced by two-photon microscopy methods for in-vivo (turbid living brains) and in-vitro (optically cleared fixated brain slices) imaging. Due to the sensitivity of such a process in terms of noise, magnification and the subject being monitored, the data can vary in scale, resolution and SNR. This strengthens the significance of cross-domain generalization capabilities in this field, and supports our cross-dataset evaluation.

All datasets are normalized by their mean and standard deviation, followed by a range stretching, such that the minimal value is 0 and the maximal is 1. 

\noindent\textbf{The DeepVess dataset} 
consists of one $200\times256\times256$ ($z\times x\times y$) in-vivo vascular image volume with expert annotations, divided into train and test. 24 additional vascular images are provided without annotation. For a fair comparison, we did not use the latter in our experiments.\\

\textbf{The VesselNN dataset} consists of 12 $~20\times256\times256$ mouse cortex and tumor vascular image volume with ground truth annotations, divided into 10 train and 2 test volumes.\\
\textbf{4D-NVIV} 
(Neurovascular Intravital Volumetric) dataset tracks neuronal activity and vascular cross-section across an entire volume of a living mouse brain. It consists of a 4D movie spanning $110\times512\times108\times67527$ voxels. As done in previous work, we segment the 3D data obtained after integrating over time. 

\smallskip
\noindent\textbf{Evaluation metrics\quad}
We evaluate our results using the following common segmentation metrics. 
Average Precision (AP) is the average precision extracted from the precision-recall graph. 
F1 score = $2 \cdot (precision \cdot recall) / (precision + recall)$. Sensitivity = $TP / (TP +FN)$, Specificity = $TN/ (TN + FP)$, TN and FP being the true positive rate and the false positive rate, respectively. Jaccard index (JI) = $TP / (TP + FP + FN)$. DICE = $(2 \cdot TP) / (2 \cdot TP + FP + FN)$. Mean Intersection Over Union (mIoU) is the mean of all JI calculated for a sequence of thresholds between 0-1.

\subsection{Results}
Tab.~\ref{tab:deepvess} and ~\ref{tab:vesselnn} present the results for the two annotated datasets, where we evaluate our results using common segmentation metrics.
As can be seen, our method outperforms all methods in the majority of metrics, where due to the tradeoff between sensitivity and specificity, the F1 score and the Average Precision are the more informative metrics. 
Our method outperform all other method in these metrics.

\noindent\textbf{Cross-Datasets evaluation\quad}
In the cross-dataset evaluation, each method is trained on dataset $A$ and evaluated on dataset $B$. As can be seen from Tab.~\ref{tab:cross}, our method significantly outperforms supervised methods. We also provide results for the fine-tuning technique, which seem to contribute most in the case of transitioning from the DeepVess dataset to the VesselNN one. Since VesselNN is more diverse and considerably larger, unsupervised fine-tuning is less effective in the other direction.

Qualitative results from both the within dataset and cross dataset experiments are shown in Fig.~\ref{fig:qual_results}. Both our method and the supervised DeepVess method present a good overlap with the ground truth data on the within dataset evaluation of the DeepVess data. On the second dataset, our results are markedly better for this protocol. On the cross dataset experiment, the gap in performance in support of our method is sizable in both datasets.

\medskip
\noindent\textbf{4D datasets segmentation\quad}
The 4D-NVIV dataset differs from previous datasets in its larger field-of-view and low SNR, due to the high acquisition rate. We ran both our method and the DeepVess method after training on the larger and more diverse dataset of VesselNN. 

\begin{figure}[t]
    \centering
    \setlength{\tabcolsep}{1pt} 
    \renewcommand{\arraystretch}{1} 
    \begin{tabular}{ccccc}
         & z=50$\mu m$ & z=100$\mu m$ & z=125$\mu m$ & z=150$\mu m$\\
        \raisebox{9mm}{\rotatebox[origin=c]{90}{\parbox{1.5cm}{\small\centering Input}}}&
        \includegraphics[width=0.23\linewidth]{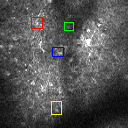} &
        \includegraphics[width=0.23\linewidth]{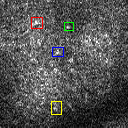} &
        \includegraphics[width=0.23\linewidth]{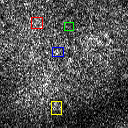} &
        \includegraphics[width=0.23\linewidth]{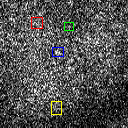} \\
        
        \raisebox{9mm}{\rotatebox[origin=c]{90}{\parbox{1.5cm}{\small\centering Ours}}}&
        \includegraphics[width=0.23\linewidth]{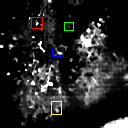} &
        \includegraphics[width=0.23\linewidth]{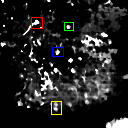} &
        \includegraphics[width=0.23\linewidth]{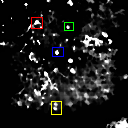} &
        \includegraphics[width=0.23\linewidth]{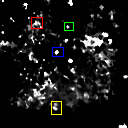} \\
        
        \raisebox{9mm}{\rotatebox[origin=c]{90}{\parbox{1.5cm}{\small\centering DeepVess}}}&
        \includegraphics[width=0.23\linewidth]{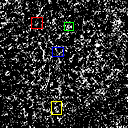} &
        \includegraphics[width=0.23\linewidth]{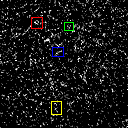} &
        \includegraphics[width=0.23\linewidth]{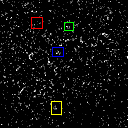} &
        \includegraphics[width=0.23\linewidth]{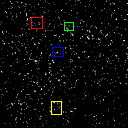}
    \end{tabular}
    \caption{Qualitative results from a 4D movie. \textbf{Columns} - different depths from brain surface are shown sequentially. The SNR decreases with imaging depth. \textbf{Rows} - Input, Our output, DeepVess output. \textbf{Colored boxes} - expert markers of penetrating vessels.}
     \label{fig:4d}
\end{figure}

The data is completely unannotated and evaluation was done by experts, who examined the output and preferred, in all cases our output over that obtained by DeepVess. To illustrate the preference, the experts have tracked penetrating vessels across the 3D stack. This was done for four blood vessels, marked by red, green, blue and yellow squares in Fig.~\ref{fig:4d}. It is evident that our method is able to produce a more reliable segmentation, where the vessel topology can be observed across cortical layers. For example, the red vessel is a penetrating artery, from which a smaller arteriole branches off at a depth of $z=100 \mu m$ below cortical surface. The ability to segment vessels in 4D data enables the highly sought after analysis of cerebral blood flow in dynamic, large field of view, imaging.

\begin{figure}[t]
     \centering
    \begin{tabular}{@{}c@{~~}c@{~~}c@{}}
        \includegraphics[width=0.312\linewidth]{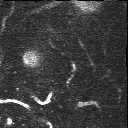} &
        \includegraphics[width=0.312\linewidth]{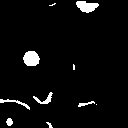} &
        \includegraphics[width=0.312\linewidth]{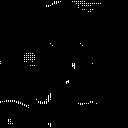} \\
        (a) & (b) & (c)
    \end{tabular}
    \caption{Illustration of the effect caused by replacing $S$ with $\bar{S}$ in Eq.~\ref{eq:loss_ac}. (a) the input image. {\color{black}(b) the output of our method.} (c) the output of our method when not employing the morphological layers.}
    \label{fig:ablation}
\end{figure}

\medskip
\noindent\textbf{Speed Performance\quad}
As can be seen from the quantitative results in Tab.~\ref{tab:deepvess}-~\ref{tab:cross}, and the qualitative results of the 4D dataset in Fig.~\ref{fig:4d}, our method outperforms the previous methods. We further look at the computational time that each method needs to process results. Because each method outputs a different output shape, we report the computation time in units of seconds per voxel. 

\begin{figure}[t]
    \centering
    \includegraphics[width=.95\linewidth]{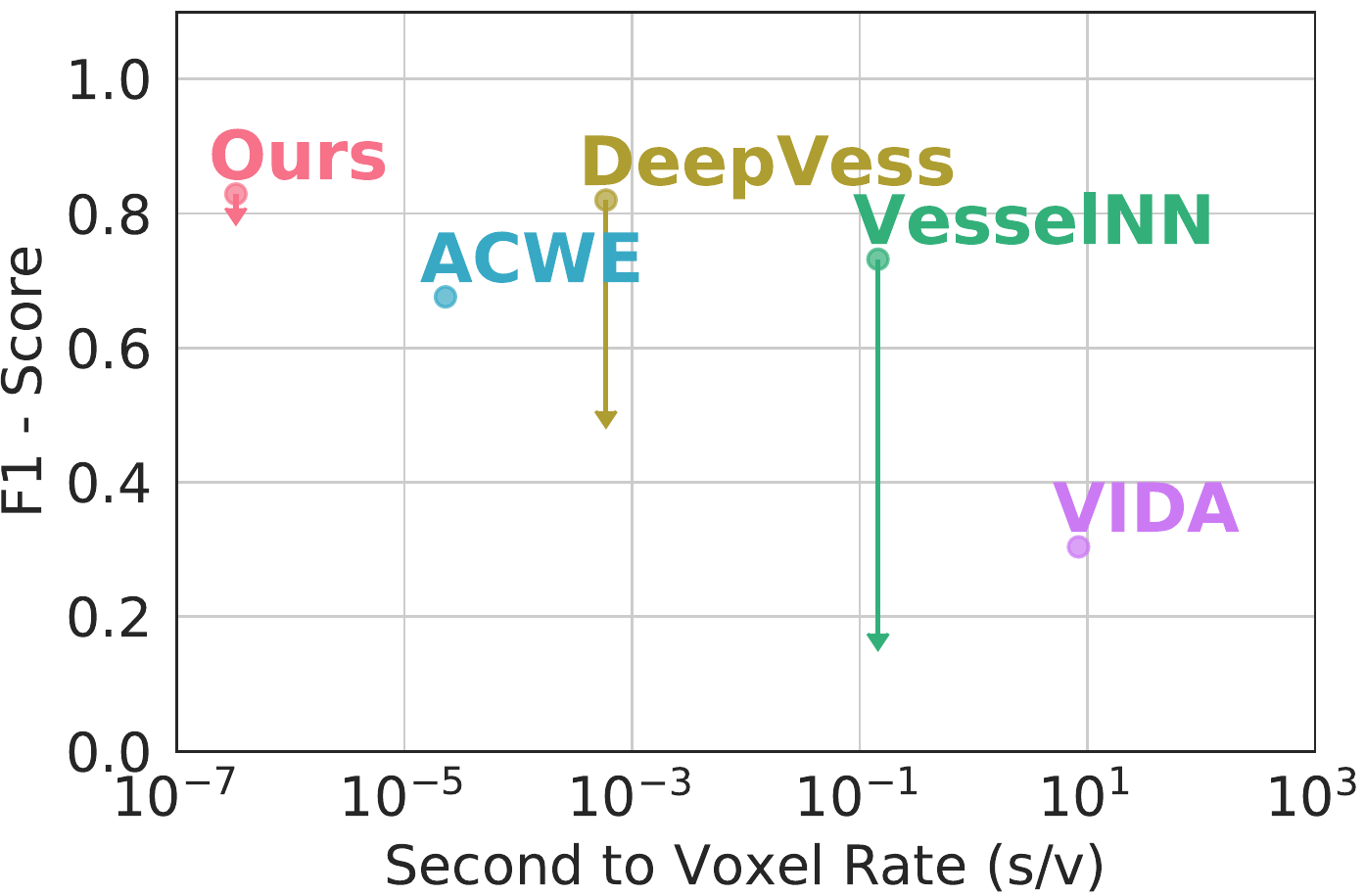}
    \caption{Computation rate vs. F1-score on DeepVess dataset. Vector lines represent the drop in performance, when performing cross-dataset evaluation.}
    \label{fig:speed}
\end{figure}

The F1-score on the DeepVess dataset is used as the other axis, and are shown for the same dataset scenario, as well as for the cross dataset training, in which the VesselNN dataset was used for training. As can be seen in Fig.~\ref{fig:speed}, our method outperforms the previous work along both axes. 

\smallskip
\noindent\textbf{Ablation Study\quad}
We examined the significance of the different losses used in our method, by removing these losses one by one. Specifically, the following variants were tests: 
(i) without $\mathcal{L}_{rank}$, in which case the disjunctive loss can collapse the output of all zero or all one. (ii) without $\mathcal{L}_{AC}$, in which case the optimization becomes a variant of adaptive thresholding. (iii) replacing $S$ with $\bar{S}$ in Eq.~\ref{eq:loss_ac}, examining the effect of the morphological pooling layer. (iv) using only $\mathcal{L}_{AC}$, which is new loss we introduce. (v) using only $\mathcal{L}_{AC}+\mathcal{L}_{rank}$, which are the two main losses.

As can be seen in Tab.~\ref{tab:ablation2}, the role of both $\mathcal{L}_{AC}$ and $\mathcal{L}_{rank}$ is very significant. One can note that the use of $\mathcal{L}_{AC}$ (Eq.~\ref{eq:loss_ac}) as a single loss already produces favorable results, supporting this novel loss term. 

The importance of the morphological layers can be observed by the drop in performance, when replacing $S$ by $\bar{S}$ in Eq.~\ref{eq:loss_ac}. Visually, as can be seen in Fig.~\ref{fig:ablation}, this results in a stripped output. We hypothesize that this pattern emerges as a way to increase the local gradient in Eq.~\ref{eq:gamma}. 

\begin{table}[t]
    \begin{tabularx}{\linewidth}{Xccccc}
        \toprule
         Losses & AP & F1 & JI & DICE & mIoU \\
         \midrule
         $\mathcal{L}$ w/o $\mathcal{L}_{rank}$ & 0.100 & 0.159 & 0.086 & 0.159 & 0.459 \\
         $\mathcal{L}$ w/o $\mathcal{L}_{AC}$ & 0.617 & 0.437 & 0.280 & 0.437 & 0.608 \\
         Eq.~\ref{eq:loss_ac} w/ $\bar{S}$ & 0.831 & 0.707 & 0.618 & 0.707 & 0.779 \\
         $\mathcal{L}_{AC}$ & 0.866 & 0.798 & 0.665 & 0.798 & 0.805 \\
         $\mathcal{L}_{AC}$+$\mathcal{L}_{rank}$ & 0.889 & 0.817 & 0.689 & 0.817 & 0.821 \\
         \midrule
         Ours $\mathcal{L}$ & \textbf{0.909} & \textbf{0.829} & \textbf{0.708} & \textbf{0.829} & \textbf{0.838} \\
         \bottomrule
    \end{tabularx}
    \caption{Ablation study showing results for various losses.}
    \label{tab:ablation2}
\end{table}

\section{Conclusion}

The quantitative analysis of blood vessel microscopy images requires accurate segmentation capabilities. While the task can be solved in a supervised way, the lack of carefully curated training data and the large domain shift between experimental settings, create the need for alternative methods.

The morphological ACWE method is a relatively recent active contour algorithm that is well-motivated and robust. In our work, it is reincarnated as a deep learning method that is able to benefit from employing a training set, as well as from the accumulated experience in our community of effectively employing CNNs. Our work, therefore, provides not just a state of the art solution that is based on novel losses and new types of layers, but also a case study for turning classical and powerful computer vision techniques into deep learning methods. 

\section*{Acknowledgement}
This project has received funding from the European Research Council (ERC) under the European Union's Horizon 2020 research and innovation programme (ERC grants CoG 725974 and StG 639416). The contribution of the first author is part of a Ph.D. thesis research conducted at Tel Aviv University. The authors thank David Kain for conducting the mouse surgery and Yulia Mitiagin for segmenting the said datasets using the VIDA suite ~\cite{Tsai2009CorrelationsON}.

{\small
\bibliographystyle{ieee}
\bibliography{main}
}

\appendix
\section{The decoder architecture}
The Decoder architecture for $D_{seg}$ and $D_{rec}$ is described in the following table:
\begin{table}[t]
    \centering
    \begin{tabular}{lcccc}
        \toprule
        Layer & In-Out & Kernel & Stride & Padding\\
        \midrule
        ConvT3D-BR & 512-64 & (4,4,4) & (2,2,2) & (1,1,1)\\
        ConvT3D-BR & 64-64 & (1,3,3) & (1,1,1) & (0,1,1)\\
        Upsample & - & - & - & - \\
        \midrule
        ConvT3D-BR* & 320-32 & (4,4,4) & (2,2,2) & (1,1,1)\\
        ConvT3D-BR & 32-32 & (1,3,3) & (1,1,1) & (0,1,1)\\
        Upsample & - & - & - & - \\
        \midrule
        ConvT3D-BR* & 160-8 & (4,4,4) & (2,2,2) & (1,1,1)\\
        ConvT3D-BR & 8-8 & (1,3,3) & (1,1,1) & (0,1,1)\\
        Upsample & - & - & - & - \\
        ConvT3D & 8-1 & (3,3,3) & (1,1,1) & (1,1,1)\\
        Sigmoid & - & - & - & - \\
        \bottomrule
    \end{tabular}
    \caption{The layers of the decoders.}
    \label{tab:decoder}
\end{table}

 We denote by ConvT3D-BR the Transposed Convolution Layer followed by Batchnorm and ReLU. The asterisk sign represents a skip connection for input $C3$ enters the 2nd block and $C2$ the third block.
 
\section{4D-NVIV results}
\begin{figure*}
    \setlength{\tabcolsep}{1pt} 
    \renewcommand{\arraystretch}{0.5} 
    \centering
    \begin{tabular}{ccc}
        & Input & Output segmentation \vspace{1mm}\\
        \raisebox{9mm}{\rotatebox[origin=c]{90}{\parbox{1.5cm}{\small\centering z=50$\mu m$}}}&
        \includegraphics[width=0.5\linewidth]{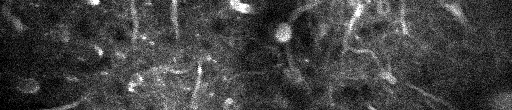} &
        \includegraphics[width=0.5\linewidth]{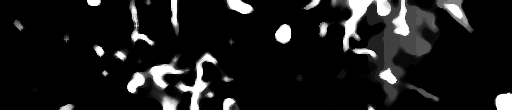} \\
        \raisebox{9mm}{\rotatebox[origin=c]{90}{\parbox{1.5cm}{\small\centering z=67$\mu m$}}}&
        \includegraphics[width=0.5\linewidth]{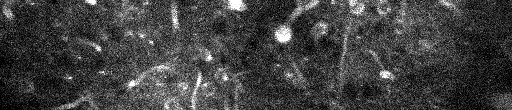} &
        \includegraphics[width=0.5\linewidth]{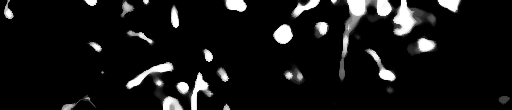} \\
        \raisebox{9mm}{\rotatebox[origin=c]{90}{\parbox{1.5cm}{\small\centering z=83$\mu m$}}}&
        \includegraphics[width=0.5\linewidth]{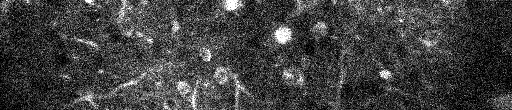} &
        \includegraphics[width=0.5\linewidth]{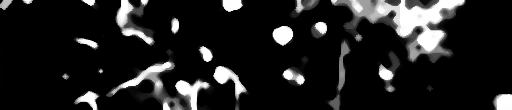} \\
        \raisebox{9mm}{\rotatebox[origin=c]{90}{\parbox{1.5cm}{\small\centering z=100$\mu m$}}}&
        \includegraphics[width=0.5\linewidth]{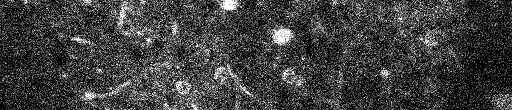} &
        \includegraphics[width=0.5\linewidth]{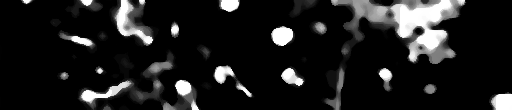} \\
        \raisebox{9mm}{\rotatebox[origin=c]{90}{\parbox{1.5cm}{\small\centering z=117$\mu m$}}}&
        \includegraphics[width=0.5\linewidth]{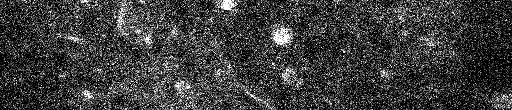} &
        \includegraphics[width=0.5\linewidth]{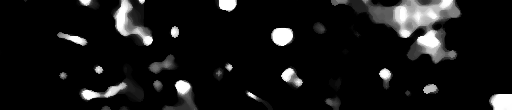} \\
        \raisebox{9mm}{\rotatebox[origin=c]{90}{\parbox{1.5cm}{\small\centering z=133$\mu m$}}}&
        \includegraphics[width=0.5\linewidth]{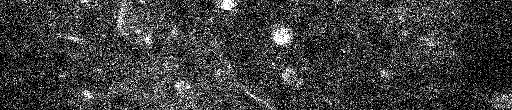} &
        \includegraphics[width=0.5\linewidth]{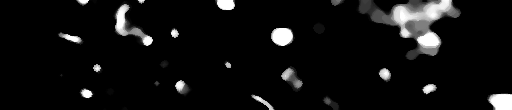} \\
        \raisebox{9mm}{\rotatebox[origin=c]{90}{\parbox{1.5cm}{\small\centering z=150$\mu m$}}}&
        \includegraphics[width=0.5\linewidth]{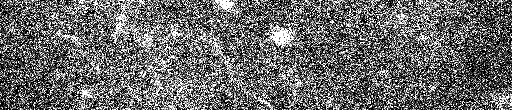} &
        \includegraphics[width=0.5\linewidth]{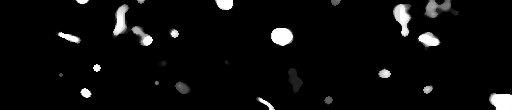} \\
        \raisebox{9mm}{\rotatebox[origin=c]{90}{\parbox{1.5cm}{\small\centering z=166$\mu m$}}}&
        \includegraphics[width=0.5\linewidth]{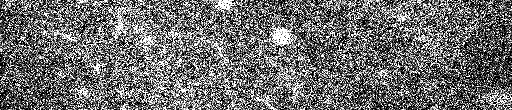} &
        \includegraphics[width=0.5\linewidth]{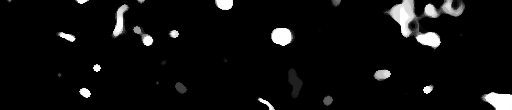} 
    \end{tabular}
    \caption{Time-collapsed $92\times440$ $ \mu m ^2$ projections of the first 4D movie at increasing depths from brain surface (left) and their respective segmentation masks (right). Neuronal somata and branching blood vessels are localized to specific axial slices whereas penetrating blood vessels travel through most axial slices. Note the robustness of the segmentation mask to degradation of the signal-to-background ratio with imaging depth.}
    \label{fig:9}
\end{figure*}

\begin{figure*}
    \setlength{\tabcolsep}{1pt} 
    \renewcommand{\arraystretch}{0.5} 
    \centering
    \begin{tabular}{ccc|ccc}
        & Input & Output segmentation & & Input & Output segmentation 
        \vspace{1mm} \\
        \raisebox{20mm}{\rotatebox[origin=c]{90}{\parbox{1.5cm}{\small\centering z=50$\mu m$}}}&
        \includegraphics[width=0.25\linewidth]{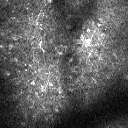} &
        \includegraphics[width=0.25\linewidth]{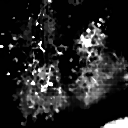} &
        \raisebox{20mm}{\rotatebox[origin=c]{90}{\parbox{1.5cm}{\small\centering z=67$\mu m$}}}&
        \includegraphics[width=0.25\linewidth]{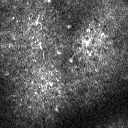} &
        \includegraphics[width=0.25\linewidth]{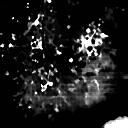} \\
        \raisebox{20mm}{\rotatebox[origin=c]{90}{\parbox{1.5cm}{\small\centering z=83$\mu m$}}}&
        \includegraphics[width=0.25\linewidth]{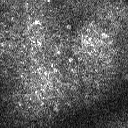} &
        \includegraphics[width=0.25\linewidth]{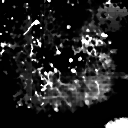} &
        \raisebox{20mm}{\rotatebox[origin=c]{90}{\parbox{1.5cm}{\small\centering z=100$\mu m$}}}&
        \includegraphics[width=0.25\linewidth]{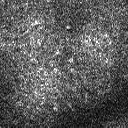} &
        \includegraphics[width=0.25\linewidth]{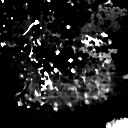} \\
        \raisebox{20mm}{\rotatebox[origin=c]{90}{\parbox{1.5cm}{\small\centering z=117$\mu m$}}}&
        \includegraphics[width=0.25\linewidth]{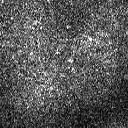} &
        \includegraphics[width=0.25\linewidth]{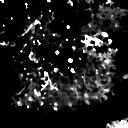} &
        \raisebox{20mm}{\rotatebox[origin=c]{90}{\parbox{1.5cm}{\small\centering z=133$\mu m$}}}&
        \includegraphics[width=0.25\linewidth]{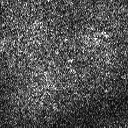} &
        \includegraphics[width=0.25\linewidth]{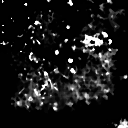} \\
        \raisebox{20mm}{\rotatebox[origin=c]{90}{\parbox{1.5cm}{\small\centering z=150$\mu m$}}}&
        \includegraphics[width=0.25\linewidth]{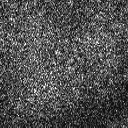} &
        \includegraphics[width=0.25\linewidth]{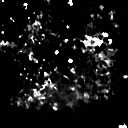} &
        \raisebox{20mm}{\rotatebox[origin=c]{90}{\parbox{1.5cm}{\small\centering z=166$\mu m$}}}&
        \includegraphics[width=0.25\linewidth]{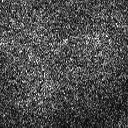} &
        \includegraphics[width=0.25\linewidth]{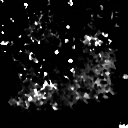}
    \end{tabular}
    \caption{Time-collapsed $430\times440$ $ \mu m ^2$ projections of the second 4D movie at increasing depths from brain surface and their respective segmentation masks. Neuronal somata and branching blood vessels are localized to specific axial slices whereas penetrating blood vessels travel through most axial slices. Note the robustness of the segmentation mask to degradation of the signal-to-background ratio with imaging depth.}
    \label{fig:28}
\end{figure*}

We present our segmentation results on two samples from the 4D intravital datasets. Please see \cite{har2018pysight} for a full description of surgical procedures, transgenic mouse lines and adherence to IACUC and local ethical guidelines. The first dataset (Fig.~\ref{fig:9}) spans $110\times512\times108\times33764$ voxels which correspond to a volume of $92\times440\times200$ $ \mu m ^3$ imaged over 268 seconds at a rate of 125.87 volumes per second. A second dataset (Fig.~\ref{fig:28}) was consecutively acquired within the same imaging session, in the same mouse, at the same magnification, and was centered on the same field of view, with the same scanning angle in x and y. It spans $110\times512\times108\times67527$ voxels which correspond to a volume of $430\times440\times200$ $ \mu m ^3$ imaged over 536 seconds at a rate of 125.87 volumes per second. Numerous neuronal cell bodies and an ensemble of penetrating vessels are visualized using genetically encoded calcium indicators and fluorescent dyes, respectively.

\clearpage
\end{document}